\documentclass[12pt,a4paper,onecolumn,tightenlines,amsmath,superscriptaddress,amssymb,nofootinbib]{revtex4}

\usepackage{natbib}
\usepackage{hyperref}
\usepackage{cleveref}

\usepackage[english]{babel}

\usepackage{enumitem}

\usepackage{graphicx}
\usepackage[font={small,it}]{caption}
\usepackage{subcaption}
\usepackage{epstopdf}

\usepackage{bm}
\usepackage{bbm}

\usepackage{amsmath}
\usepackage{amsthm}
\usepackage{amssymb}
\usepackage{xfrac}

\usepackage{lscape}
\usepackage{tabularx}
\usepackage{gensymb}

\usepackage{enumitem}

\newcounter{tlc}

\crefname{tlc}{}{}

\newtheorem*{theorem*}{Theorem}
\newtheorem*{lemma*}{Lemma}
\newtheorem*{corollary*}{Corollary}

\usepackage{enumitem}
\allowdisplaybreaks[1] 
\usepackage[super]{nth}

\newcommand{\ket}[1]{|#1\rangle}
\newcommand{\braket}[2]{\langle #1|#2\rangle}
\newcommand{\ketbra}[2]{|#1\rangle\!\langle#2|}
\newcommand{\id}{\mathbbm{1}}
\newcommand{\expt}[1]{\left\langle #1\right\rangle}
\DeclareMathOperator{\tr}{Tr}
\DeclareMathOperator{\diag}{diag}

\makeatletter
\DeclareRobustCommand{\cev}[1]{%
  \mathpalette\do@cev{#1}%
}
\newcommand{\do@cev}[2]{%
  \fix@cev{#1}{+}%
  \reflectbox{$\m@th#1\vec{\reflectbox{$\fix@cev{#1}{-}\m@th#1#2\fix@cev{#1}{+}$}}$}%
  \fix@cev{#1}{-}%
}
\newcommand{\fix@cev}[2]{%
  \ifx#1\displaystyle
    \mkern#23mu
  \else
    \ifx#1\textstyle
      \mkern#23mu
    \else
      \ifx#1\scriptstyle
        \mkern#22mu
      \else
        \mkern#22mu
      \fi
    \fi
  \fi
}
\makeatother

\usepackage{stackengine}

\stackMath

\newcommand{\kB}{k_\mathrm{B}}

\newcommand{\KL}{Kullback--Leibler}

\newcommand{\Ent}[1]{{H}\small(#1\small)}

\newcommand{\rDiv}[3]{D_{#1}\!\left(#2 \,||\,#3 \right)}
\DeclareMathOperator{\supp}{support}

\usepackage{color}
\usepackage[dvipsnames]{xcolor}
\usepackage[normalem]{ulem}

\newcommand{\CQT}{Centre~for~Quantum~Technologies, National~University~of~Singapore, 3 Science Drive 2, 117543, Singapore} 
\newcommand{\IQOQI}{Institute for Quantum Optics and Quantum Information,\\
Austrian Academy of Sciences, Boltzmanngasse 3, A-1090 Vienna, Austria}

\begin{document} 

\title{One-shot information-theoretical approaches \\ to \mbox{fluctuation theorems}}
\author{Andrew J. P. Garner}
\email{physics@ajpgarner.co.uk}
\address{\CQT}
\address{\IQOQI}

\date{\today}

\begin{abstract}
\hyphenpenalty=10000
Traditional thermodynamics governs the behaviour of large systems that evolve between states of thermal equilibrium.
For these large systems, the {\em mean} values of thermodynamic quantities (such as work, heat and entropy) provide a good {characterisation} of the process.
Conversely, there is ever-increasing interest in the thermal behaviour of systems that evolve quickly and far from equilibrium,
 and that are too small for their behaviour to be well-described by mean values.
Two {major} fields of modern thermodynamics seek to tackle such systems:
 \mbox{\em non-equilibrium thermodynamics},
 and the nascent field of {\em one-shot statistical mechanics}.
The former provides tools such as {\em fluctuation theorems}, whereas the latter applies {\em ``one-shot'' R\'enyi entropies} to thermal contexts.
In this chapter of the upcoming book ``Thermodynamics in the quantum regime -- Recent progress and outlook'' (Springer International Publishing), I provide a gentle {introduction} to recent research that draws from both fields: the application of one-shot {information} theory to fluctuation theorems.
\end{abstract}

\maketitle

\enlargethispage{\baselineskip}
Modern technological developments have driven interest in the thermal properties of systems of ever-diminishing size.
The build-up of dissipated heat is a limiting factor on the speed of microprocessors.
Over four decades of adherence to Moore's law~\cite{Moore65} (an exponential decrease in the size of electronics) has shrunk the size of transistors on commercial chips to $10\;{\rm nm}$~(e.g.\ \cite{Intel17}),
 and there are experimental demonstrations of transistors as small as just $7$ atoms~\cite{Fuechsle10}.
The advent of quantum computing~\cite{NielsenC00} takes computation to an even smaller scale, where the fundamental unit of quantum information~-- a qubit -- may be physically represented by a choice between two energy levels of a {\em single} atom~\cite{CiracZ95,Ladd10}.
As such, there is a pressing need to understand and characterize the thermal behaviour of extremely small systems.

On the other hand, the traditional laws of thermodynamics~\cite{BlundellB06} are understood to govern the behaviour of asymptotically large ensembles of independent systems.
Here, by the law of large numbers, properties of the system observed in any given experimental run closely match the average value.
Moreover, traditional thermodynamics applies to processes where one compares states of a system in thermodynamic equilibrium---completely characterized by a few state variables, such as temperature or free energy.
Doing this implies an additional assumption on the system: namely, that it must spend a sufficiently long (theoretically, sometimes infinite) amount of time thermalizing during any process. 

To what extent does it make sense to use thermal quantities, such as {\em heat} and {\em work}, outside of these large, slow settings?
Two major fields within modern thermodynamics seek to address this.
The first field, {\em non-equilibrium statistical mechanics} (see e.g.\ \cite{ZubarevMR96}), 
 characterizes the behaviour of systems taken out of thermal equilibrium.
A notable approach that I shall discuss in this chapter are {\em fluctuation relations}~\cite{Jarzynski97,Crooks98,Crooks99}, 
 which relate the non-equilibrium behaviour of a driven system to equilibrium values such as free energy.
The second field {\em one-shot statistical mechanics} (see e.g.\ \cite{DelRioARDV11,DahlstenRRV11,Dahlsten13,Aberg13,HorodeckiO13,EgloffDRV15,BrandaoHNOW15,WeilenmannKFR16}) 
 draws techniques from one-shot information theory~\cite{Renyi61,RennerW04,Renner05,Tomamichel16} to tackle deviations in statistical behaviour
 when the mean regime no longer describes a process well.

In this chapter, I discuss a modern approach that draws from both fields: the application of one-shot information theoretic quantities to characterize processes governed by fluctuation relations.
I aim here to provide a gentle introduction to the topic, rather than a comprehensive review,
 and as such shall spend a fair amount of time explaining the prerequisite concepts.
I begin in \cref{sec:Entropy} with a review of information-theoretic entropy.
Here I will attempt to provide some intuition for one-shot entropies, relative entropies (divergences), and the quantum extensions of these quantities.
In \cref{sec:ThermoTools}, I outline the role of information within thermodynamics, and then present a generic setting in which we can discuss both one-shot and non-equilibrium thermodynamics.
In \cref{sec:WorkFluctuations}, I discuss how one-shot entropies fit with the formalism of {\em fluctuation theorems},
 before finally in \cref{sec:QuantumFluctuations}, I consider the case when those fluctuation theorems are {\em quantum}.

\section{One-shot entropies and divergences}
\label{sec:Entropy}

\subsection{The Shannon and R\'enyi entropies}
\label{sec:RenyiEntropy}
Consider %
 a random variable $X$ that corresponds to choice of $x_i$ from alphabet $\mathcal{X}$, where $x_i$ is selected with probability $p_x(i)$.
The {\em Shannon entropy} $H(X)$ is then defined~\cite{Shannon48}:
\begin{equation}
\label{eq:Shannon}
\Ent{X} = -\sum_i p_x(i) \ln p_x(i).
\end{equation}
It is common to refer to $\Ent{X}$ as the {\em information} of random variable $X$.
Although one can take a formal axiomatic approach to deriving this quantity~\cite{Shannon48,Renyi61,HanelT13},
 let us here present a looser intuitive understanding~\cite{Jones79,Vedral02Rev}:
One can consider $\ln\left[\sfrac{1}{p_x(i)}\right]$ as a measure of ``surprise'' on receiving some output $x_i$. 
The smaller the $p_x(i)$, the greater the surprise\footnote{The surprise involves a logarithm rather than just the reciprocal probability to ensure additivity of surprises. 
For two mutually exclusive events with joint probability $p\cdot q$, the joint surprise $-\ln\left(p\cdot q\right) = -\ln\left(p\right) -\ln\left(q\right)$.}.
In this picture, the Shannon entropy is therefore the {\em average surprise} one experiences upon sampling $X$, 
 since the various surprises of each outcome are weighted by $p_x(i)$ in the sum Eq.~\eqref{eq:Shannon}.

Information theory is typically concerned about asymptotic limits.
We must therefore exercise caution when using its tools to describe the behaviour of small systems, 
 or the statistics formed through limited repetitions of an experiment. 
Let us reflect on the physical meaning of the Shannon entropy, by considering how information is {\em encoded} onto a physical system.
Suppose we wish to encode random variable $X$ onto a physical system $\Xi$.
The na\"ive temptation is to assign one microstate of the system for each variate $x_i$ in $\mathcal{X}$.
However, the behaviour of the system under this encoding will not match that predicted by equations involving the Shannon entropy,
 except in the special case of distributions where each variate $x_i$ occurs with equal probability.
Rather, the meaning of Shannon entropy comes from the {\em source coding theorem}~\cite{Shannon48}:
 that in the limit of large $N$, $N$ independently and identically distributed ({\em i.i.d.}) instances of a random variable $X$ can be encoded onto a physical medium with $e^{N \Ent{X}}$ configurations\footnote{The base here is $e$ because we have defined $H$ in units of {\em nats}. 
If we had defined $H$ in {\em bits} using $-\sum_i p_x(i) \log_2 p_x(i)$, the base would be $2$.}.
In other words, in the limit of large $N$, there is always an optimal encoding that allows one to store (or to send) strings of length $N$ in a physical system with $e^{N \Ent{X}}$ configurations, and to recover the exact string encoded almost perfectly (i.e.\ with arbitrarily small error probability).
Conversely, if the physical medium has fewer than $e^{N \Ent{X}}$ configurations, then there will almost always be messages that cannot be recovered.

The existence of such an encoding follows from the principle of {\em typical sequences}, which in turn relates to the law of large numbers---valid, as implied by the name, only when $N$ is sufficiently large.
The set of typical sequences\footnote{
In the context of i.i.d.\ $X$, a sequence will be typical if and only if each symbol $x_i$ appears $p_i N$ times.
} has size $e^{N H(X)}$, as opposed to the much greater number $|\mathcal{X}|^N$ of total possible strings.
Moreover, every typical sequence will occur with equal probability.
For large $N$, the probability that a randomly sampled string is a {\em typical sequence} approaches unity.
In this asymptotic limit, the efficient encoding is then to associate each configuration of the physical system with a typical sequence. 

This encoding, however, is strictly asymptotic.
Consider the distribution 
\begin{equation}
\label{eq:ExampleDist}
P(X) = \left[\frac{1}{2},\; \frac{1}{4},\; \frac{1}{8},\; \frac{1}{16},\; \frac{1}{16}\right]
\qquad
\mathrm{over}
\qquad
\mathcal{X} = \{x_a,\; x_b,\;  x_c,\;  x_d,\;  x_e\}.
\end{equation}
This has a Shannon entropy of $\frac{15}{8} \ln 2 \;\mathrm{nats}$,
 so by the source coding theorem there is an asymptotic encoding for strings of length $N$ that requires $e^{N \frac{15 \ln 2 }{8}}$ distinct configurations. 
On the other hand, suppose we only have one copy of $X$ that we wish to encode onto a single physical system.
Na\"ively applying the source coding theorem suggests that we only require $\approx 3.87$ configurations,
 but even rounding up to $4$, it is immediately obvious that a physical system with $4$ configurations is insufficient to reliably encode a single message:
 the best (most likely to succeed) scheme is one where we allocate each of $x_a$, $x_b$ and $x_c$, and one of $x_d$ or $x_e$ to the $4$ physical configurations.
Here, there remains a probability $\frac{1}{16}$ of failing to encode the message\footnote{Or conversely, we encode both $x_d$ and $x_e$ to the same physical configuration. 
Then, $\frac{1}{8}$ of messages have an indistinguishable $x_d$/$x_e$, and we can guess the correct symbol $\frac{1}{2}$ of the time: resulting in a $\frac{1}{16}$ probability that the message is incorrectly decoded.}%
---much worse than the certainty of success promised by the source encoding theorem.
Indeed, we see that when $N=1$, five distinct configurations are necessary to encode a single message $X$ without error: one for each possible symbol in the set $\mathcal{X}$.

This motivates the search for additional information quantities beyond the Shannon entropy when dealing with small $N$.
In the above example, the pertinent quantity is in fact the {\em max-entropy} (or {\em Hartley entropy}) $H_0$, which is defined
\begin{equation}
\label{eq:MaxEntropy}
H_0:= \ln \left| \supp \left(X\right) \right|
\end{equation}
(where the {\em support} is the set of elements in $\mathcal{X}$ that occur with probability strictly greater than $0$, 
 and $|\supp\left( X\right)|$ is the number of these nonzero elements).
In our example, $H_0(X) = \ln 5$, and it obvious that a physical system with $e^{\ln 5} =5$ configurations would be sufficient to reliably encode a single instance of $X$.
One alternative intuition of $H_0(X)$ is that it is the Shannon entropy of the most random distribution possible that has the same support as $X$---the uniform distribution of the same size as $X$.
As such, $H_0$ always upper bounds the Shannon entropy.

More generally, the Shannon entropy and the max-entropy are both part of a broader family of {\em R\'enyi entropies}, defined~\cite{Renyi61}: 
\begin{equation}
\label{eq:RenyiEntropy}
H_\alpha(X) := \frac{1}{1-\alpha} \ln \left[ \sum_i {p_X(i)}^\alpha\right], \qquad \alpha \geq 0.
\end{equation}
We see that the special case $\alpha = 0$ corresponds to the max-entropy\footnote{
Sometimes the $H_\frac{1}{2}$ is also referred to as the max-entropy. 
To avoid ambiguity, here we label the entropy explicitly by the parameter $\alpha$, i.e.\ writing $H_0$ rather than the ambiguous ``$H_{\rm max}$''.
}; and if we take the limit $\alpha\to1$, we recover the Shannon entropy.
In general, R\'enyi entropies are non-increasing with $\alpha$, such that the {\em max-entropy} ($\alpha=0$) is the largest entropy value.
(Even broader generalizations of entropy are possible---see, for example~\cite{RennerW04} or \cite{HanelT13}.)

Let us discuss one other important R\'enyi entropy: the {\em min-entropy}, defined as $\alpha\to\infty$:
\begin{equation}
\label{eq:MinEntropy}
H_\infty(X) := \lim_{\alpha\to\infty} H_\alpha(X) = -\ln\left[ \max p_i \right].
\end{equation}
Intuitively, this quantifies the {\em worst-case randomness} of a variable in cryptographic contexts where randomness is a resource~\cite{Renner05} (here, ``worst'' means least random, as opposed to thermodynamic contexts, where states of low entropy are often more useful).
This is because $H_\infty$ quantifies the maximum amount of {\em uniform randomness} that we can extract from a string of $N$ copies of $X$, if $X$ is our only source of randomness.
In such a context, we cannot do probabilistic post-processing,
 but rather the best we may do is to {\em coarse-grain} the various strings in $\mathcal{X}^{\otimes N}$ such that the binned distribution $Y$ is uniformly random (that is apply a many-to-one map of $x^{\otimes N}$ onto $y\in\mathcal{Y}$).
In this setting, the {\em most likely} string of length $N$ will occur with probability ${p_{\rm max}} ^ N$, where $p_{\rm max} := \max_i p_x(i)$.
Since all strings from $X^{\otimes N}$ must be included in a bin, 
 and inclusion of additional strings in the bin can only ever increase the total probability of sampling that bin,
 we see that at least one variate $y'$ of $Y$ must have at least the probability ${p_{\rm max}} ^ N$ of occurring.
Since we want a uniformly random $Y$,
 {\em at best} $\mathcal{Y}$ can then contain only ${p_{\rm max}}^{-N}$ outcomes, 
 each occurring with probability ${p_{\rm max}}^N$.
The distribution over this binned $Y$ has a (Shannon) entropy of $N \log p_{\rm max}$, which we normalize by $N$ to $\log p_{\rm max}$ per sample of $X$.

R\'enyi entropies are often colloquially referred to as ``one-shot'' quantities, 
 but the above context shows that this does not preclude them from also having asymptotic meanings.
A more directly ``one-shot intuition'' of $H_\infty$ is that it is the {\em least surprised} we can be upon learning an outcome of $X$.
In our above example [\cref{eq:ExampleDist}]: $H_{\infty} = \ln 2\;{\rm nats}$ (i.e.\ $1$ bit).
This corresponds to a randomness-extraction scheme where we coarse-grain $X$ into two sets $\{\{x_a\},\{x_b, x_c, x_d,x_e\}\}$, each occurring with probability $\frac{1}{2}$.
Alternatively, we are the least surprised by $X$ when it generates outcome $x_a$, which happens with probability $\frac{1}{2}$.

\begin{figure}[htb]
\begin{center}
\includegraphics[width=0.4\textwidth]{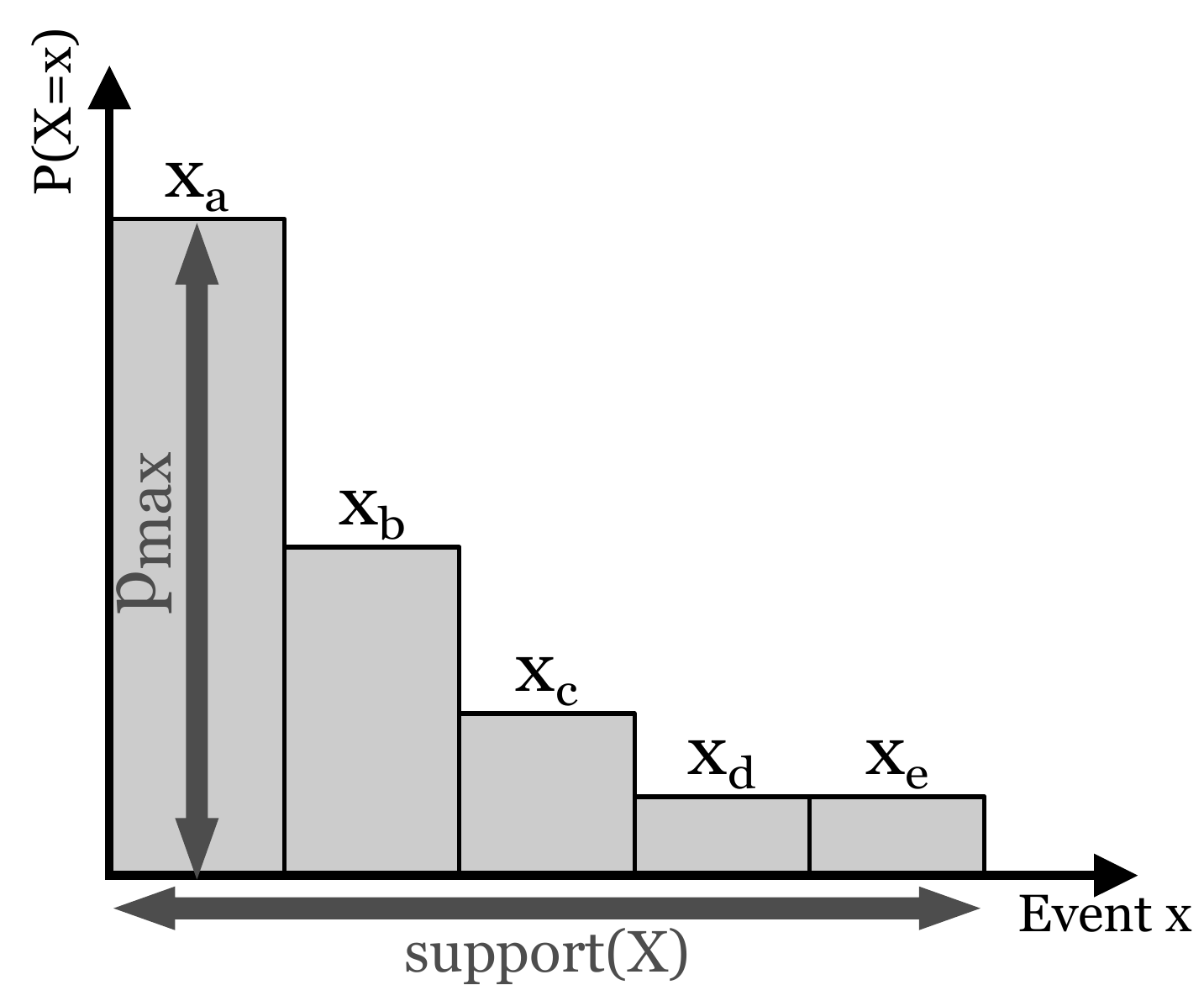}
\caption[Example probability distribution]{
\textbf{Example distribution.} 
The probability distribution $\left(\frac{1}{2}, \frac{1}{4}, \frac{1}{8},\frac{1}{16},\frac{1}{16}\right)$ of our example random variable $X$ is drawn.
The ``width'' of this distribution is the support (number of elements), $5$, and determines the max-entropy $H_0$.
The ``height'' of this distribution is the maximum probability $p_{\rm max} = \frac{1}{2}$ and determines the min-entropy $H_\infty$.
The Shannon entropy $H_1$ is a weighted function of both width and height (see also \cite{Dahlsten13}).
}
\label{fig:ExampleP}
\end{center}
\end{figure}

Different R\'enyi entropies characterize different aspects of a probability distribution (see \cref{fig:ExampleP}).
One can interpret~\cite{Dahlsten13} the parameter $\alpha$ as weighting the relative importance between the ``width'' (i.e.\ support), and the ``heights'' (read: probabilities) of the distribution.
Finally, although it is beyond the scope of this chapter, we note that R\'enyi entropies can be {\em smoothed}~\cite{RennerW04,Renner05,Tomamichel16}. 
This amounts to a reshuffling of the probability distribution (bounded according to some distance measure) in order to extremize the entropic quantity at hand. 
For instance, one could smooth $H_0(X)$ by discounting the least likely $x_i$, up to some total probability $\epsilon$.
A pedagogical introduction to this may be found in \mbox{\citet{Dahlsten13}}.

\subsection{Differential entropies}
\label{sec:DiffEntropy}
The probabilities of continuous variables are often expressed in terms of a {\em probability density function} $P(x)$, which must be integrated over a range in order to yield the probability of observing a sample within that range.
When discussing information entropies of a continuous distribution, it is often useful to use a {\em differential entropy}, since the usual information entropies are typically divergent when faced with a set of infinitesimal probabilities.
Suppose we binned $P(x)$ into bins of size $\delta x$, such that (in the limit of small $\delta x$) the probability of each bin is $P(x) \delta x$.
Then as $\delta x \to 0$, the Shannon entropy will contain divergent terms of the form $-\ln\left[P(x)\delta x\right]\to\infty$.
This is not incorrect! 
We would indeed be infinitely surprised to see any {\em exact} outcome $x$---any countable set of a real numbers is a measure zero set of a continuum.

This does not mean we have to give up on quantifying entropy for continuous distributions, though we must concede that they are, in general, infinitely more entropic than discrete ones (consider the number of bits a classical digital computer would require to perfectly store a real number to arbitrarily high accuracy~\cite{GarnerLTVG17}). 
However, by expanding $\ln\left[P(x)dx\right] = \ln\left[P(x)\right]+\ln dx$, we see it is in fact the second term that is upsetting, and this term is independent of $P(x)$.
Thus, one takes instead a {\em differential entropy}, which measures how much more entropic $P(x)$ is with respect to some implicit reference distribution. 
Moreover, this difference can be defined in terms of $dx$, where terms in $\ln dx$ cancel {\em before} the limit $dx\to0$ is taken.
This allows us to effectively ``renormalize'' the entropy.

Typically, the reference distribution is a uniform distribution of unit width and height.
This is an arbitrary choice (e.g.\ thinner uniform distributions will have negative entropies, and this has no special physical meaning).
This raises an important concern when we consider a probability density over a dimensionful quantity such as work (energy), or height (distance).
If we always choose a width of $1$ and a height as $1$ as our reference distribution then either:
 {\bf (i)} this makes no dimensional sense, 
 or {\bf (ii)} if we add units to fix the problem i (i.e.\ for a probability distribution over a quantity measured in units of $k$, take a reference distribution of width $1\;\rm{k}$ and height $1\; \rm{k}^{-1}$), then probability densities referring to the {\em same data} expressed in different units will have different differential entropies, because the implied reference distributions will be different.
For example, consider a probability distribution over heights of people, measured both in feet and in meters: the implied reference distribution for the former is a uniform distribution over $1\;{\rm foot}$, the latter over $1\;{\rm meter}$ -- and hence the differential entropy for this same physical distribution expressed in two different units will be different.

To avoid this problem, we should explicitly write our choice of reference when we define the differential entropy.
Let us take as such a reference a uniform distribution $K(x)$,
 and compare the difference in Shannon entropy between $P_{\rm bin}(x)$ and $K_{\rm bin}(x)$ -- the distributions formed by binning $P(x)$ and $K(x)$ into a discrete distributions the binned alphabet $\mathcal{B}$, and probabilities given by $P(x)\delta x$ and $K(x)\delta x$ respectively (for small $\delta x$):
\begin{align}
H_1[P_{\rm bin}(x)] - H_1[K_{\rm bin}(x)] \hspace{-5em} & \nonumber \\
& = -\sum_{x\in\mathcal{B}} P(x)\delta x \ln \left[P(x)\delta x\right] + \sum_{x\in\mathcal{B}} K(x)\delta x \ln \left[K(x)\delta x\right] \nonumber \\
& = -\sum_{x\in\mathcal{B}} \left\{ P(x)\delta x \ln \left[P(x)\right] + P(x)\delta x \ln\left[\delta x\right] \right\}
\nonumber \\
& \hspace{2em}
  + \sum_{x\in\mathcal{B}} \left\{ K(x)\delta x \ln \left[K(x)\right] + K(x)\delta x \ln \left[dx\right] \right\} \nonumber \\
& = -\sum_{x\in\mathcal{B}} P(x)\delta x \ln \left[P(x)\right] - \ln \delta x + \sum_{x\in\mathcal{B}} K(x)\delta x \ln \left[K(x)\right] + \ln \delta x \nonumber \\
& = -\sum_{x\in\mathcal{B}} P(x)\delta x \ln \left[P(x)\right] + \sum_{x\in\mathcal{B}} K(x)\delta x \ln \left[K(x)\right] \label{eq:binlimit}
\end{align}
(We have used that $\sum P(x) \delta x = \sum K(x) \delta x = 1$, such that the terms in $\ln \delta x$ exactly cancel.)
In this form, it can be explicitly seen that the difference between the entropy of two distributions does not contain any terms that will diverge as we take the limit $\delta x \to 0$, whereby the sums convert into integrals.

Suppose we choose $K(x)$ to be a uniform distribution over width $k$ with probability density $k^{-1}$, 
 then the right-most term of Eq.~\eqref{eq:binlimit} becomes $- \ln k$ [in the same units as the left-hand term in $P(x)$],
 and so the differential Shannon entropy of probability density function $P(x)$ with respect to this reference may be written
\begin{align}
H_1^k[P(x)] &:= \lim_{\delta x\to0} \left\{H_1[P_{\rm bin}(x)] - H_1[K_{\rm bin}(x)]\right\} \nonumber \\
& = -\int_{-\infty}^{\infty} P(x) \ln\left[ k \; P(x)\right] dx.
\end{align}
Similar forms exist for other differential R\'enyi entropies, for example, $H_{0}^k$ and $H_{\infty}^k$:
\begin{align}
H_0^k[P(x)] &:= \ln\left[\frac{1}{k} \supp\left[P(x)\right]\right] \\
H_\infty^k[P(x)] & := -\ln \left[k \max_x \left[P(x)\right] \right].
\end{align}
The former compares the support of $P(x)$ with the range $k$; the latter compares the maximum density of $P(x)$ with $k^{-1}$.
One additional benefit of these definitions including $k$ is that it avoids equations where one seemingly must take the logarithm of dimensionful quantities.

The values given by different choices of $k$ are related by a constant offset given by the difference in the entropy between the different implied reference distributions.
In thermodynamic settings, where we are commonly concerned with distributions over energies (work and heat),
 a natural reference is to set $k= {\kB T}$, since this is the characteristic energy scale of fluctuations at temperature $T$~\cite{YungerHalpernGDV15}.
This is formally equivalent to taking probability distributions over energies expressed in units of $\kB T$, and then using the traditional form of differential entropies where the reference distribution is implicit.

\subsection{Relative entropies and divergences}
Conceptually related to information entropies are {\em relative entropies}, also known as \mbox{\em divergences}.
While information entropies quantify the randomness intrinsic to a variable,
 the relative entropy quantifies deviations in statistical behaviour between two probability distributions.

Consider the random variables $X$ and $Y$ with probabilities of outcomes $\{p_x(i)\}$ and $\{p_y(i)\}$ respectively (wherein $X$ and $Y$ have alphabets such that the index $i$ labels the same symbol from alphabet $\mathcal{A}$).
The {\em \KL{} divergence} (or {\em average relative entropy}) is defined:
\begin{equation}
\label{eq:KullbackLeibler}
\rDiv{1}{X}{Y} = - \sum_i p_x(i) \log \dfrac{p_y(i)}{p_x(i)}.
\end{equation}

We can understand this divergence as a measure of relative surprise~\cite{Vedral02Rev}.
Suppose that we had some system  configured according to random variable $X$, but we thought it was configured according to $Y$.
Our surprise at each outcome will be given by terms of the form $\ln p_y(i)$, but these surprises occur with probabilities given by the true probabilities $p_x(i)$. 
On average our misidentified system will surprise us by $-\sum_i p_x (i) \ln p_y(i)$.
The average discrepancy between this, and the true amount ($-\ln p_x(i)$) that we should have been surprised by,
 is then \KL{} divergence $D_1$.

Equivalently, $D_1$ quantifies the average amount of information we gain when we misidentify a system $\xi$ configured in $X$ as one configured in $Y$, and then are subsequently corrected~\cite{CoverT06}.
The interpretation of \KL{} divergence as {\em information gain} gives it a role in {\em machine learning}~\cite{Mitchell97}.
Suppose a machine is attempting to learn some the distribution of system $\xi$ that is configured according to $X$.
If $Y$ is the machine's current estimate of $\xi$, upon receiving a new sample from the true distribution $X$, the average information gained is $\rDiv{1}{X}{Y}$.
(When the learning is eventually complete, $Y=X$, and $\rDiv{1}{X}{Y}=0$; here, the machine no longer learns new information upon sampling $\xi$, since its model is already perfect).

Like the Shannon entropy, $D_1$ is an asymptotic quantity, involving averages, 
 and so might have limited meaning in a regime where the number of experimental trials are limited.
Thus, as with the Shannon entropy, we can generalize this quantity to a family of R\'enyi divergences~\cite{vanErvenH14}, 
 which characterise various aspects of the discrepancy between two distributions beyond the asymptotic limit:
\begin{equation}
\rDiv{\alpha}{X}{Y} := \frac{1}{\alpha-1}\ln \left(\sum_i \frac{{p_i}^\alpha}{{q_i}^{\alpha-1}} \right), \qquad \alpha\geq 0.
\end{equation}

The limiting case $\alpha\to1$ yields the \KL{} divergence.
There are two other special cases we are particularly interested in.
The first, $\alpha=0$, quantifies a difference in support between $X$ and $Y$:
\begin{equation}
\rDiv{0}{X}{Y} = -\ln\left[ \sum_{i \in \{ i | p_x(i)>0\}} p_y(i) \right].
\end{equation}
That is, we measure a function of the probability that a sample of $Y$ is within the support of $X$.
This can be also be interpreted in a one-shot information gain context 
 but where we only consider if events are possible or not, and put no weighting on the probability of events distributed according to $X$.
If the support of $X$ and $Y$ are the same, then no single outcome of $X$ will challenge our belief that the system is configured according to $Y$--reflected by $\rDiv{0}{X}{Y}=0$.
On the other hand, if some of the outcomes in $Y$ are not in $X$, these outcomes will never be observed by sampling $\xi$,
 and so each sample makes us more likely to believe the system is not configured according to $Y$, $\rDiv{0}{X}{Y}>0$.
Finally, if there is an outcome $j$ in $X$ but not in $Y$ (where $p_x(j)>0$ but $p_y(j)=0$) then observing this outcome will make us {\em certain} that we were not configured according $X$.
Here, $\rDiv{0}{X}{Y}$ is infinite, since $j$ would be infinitely surprising to us with our erroneous distribution $Y$.

The final special case we shall discuss is $\alpha\to\infty$:
\begin{equation}
\rDiv{\infty}{X}{Y} =  -\log\left[\max_i\left(\frac{p_y(i)}{p_x(i)}\right)\right]. \label{eq:DInfty}
\end{equation}
This quantifies the most shocking discrepancy between $X$ and $Y$,
 quantifying the {\em greatest extent} we can be surprised by an outcome if we thought we had $Y$ but actually had $X$.
This does not factor in that the variate that maximises this quantity may be very unlikely to occur.
As before, if the $i^{\rm th}$ outcome occurs with some probability in $X$ but not in $Y$, then $\frac{p_y(i)}{p_x(i)} = 0$ and Eq.~\eqref{eq:DInfty} diverges:
 here, such an outcome would completely falsify our belief that we had distribution $Y$ in a single-shot.

$D_\alpha$ also obeys a monotonic relationship with $\alpha$, but in the opposite direction to the R\'enyi entropies.
$D_\alpha$ is monotonically non-decreasing with $\alpha$ such that $D_0$ will be the lowest value, and $D_\infty$ the highest.

Divergences readily generalize to continuous distributions.
Since the divergence implicitly involves a difference in the first place, the problems of renormalization intrinsic to the differential entropy (see discussion in \cref{sec:DiffEntropy}) do not appear here; no implicit reference distribution is required, since we are already directly comparing two distributions.
For two probability density functions $P(x)$ and $Q(x)$ over $x$, we have:
\begingroup
\allowdisplaybreaks
\begin{align}
\rDiv{\alpha}{P(x)}{Q(x)} & = \dfrac{1}{\alpha-1} \int_{-\infty}^{\infty} \hspace{-0.5em} dx \; \ln\left[{P(x)}^\alpha {Q(x)}^{1-\alpha}\right] \label{eq:crdAlpha}, \\
\rDiv{0}{P(x)}{Q(x)} & = 
-\ln \left( \int_{-\infty}^{\infty} dx \; \theta\!\left[P(x)\right] Q(x) \right)
 \label{eq:crd0}, \\
\rDiv{1}{P(x)}{Q(x)} & = \int_{-\infty}^{\infty} \hspace{-0.5em} dx \; P(x) \ln \frac{P(x)}{Q(x)} \label{eq:crd1}, \\
\rDiv{\infty}{P(x)}{Q(x)} & = \begin{cases}
 \ln \left(\min \{ \lambda \;|\; P(x)\leq \lambda Q(x) \; \forall x \} \right) & \supp Q \subseteq \supp X, \\
\infty & \mathrm{otherwise},
\end{cases}
 \label{eq:crdInfty}
\end{align}%
\endgroup
where in Eq.~\eqref{eq:crd0}, $\theta(y) = 1$ if $y>0$ and $0$ otherwise.

\subsection{Entropies in quantum information theory}
All of the above entropic quantities have quantum equivalents.
Classical probability distributions may be thought of as special cases of quantum states
 that are diagonal in the basis of the measurement distinguishing between the variates. 
Consider the set of variates $\mathcal{X}= \{x_i\}$ that form the alphabet of some random variable.
One may express the choice from this set as a Hilbert space of dimension $|\mathcal{X}|$.
Then for a random variable $X$ where each outcome $x_i$ occurs with probability $p_x(i)$,
 one can construct a quantum state $\rho_X = \sum_i p_x(i) \ketbra{x_i}{x_i}$ where $\{\ket{x_i}\}$ are an orthonormal basis.
The state $\rho_x$ hence has eigenvalues that are the probabilities associated with $X$, and eigenvectors that are in one-to-one correspondence with each variate $x_i$ of $X$,
 and hence knowing the matrix $\rho_X$ would allow us to reconstruct random variable $X$ if we are also given the alphabet $\mathcal{X}$.

It is natural therefore to seek quantum analogues of classical entropies that are functions of quantum states.
The further subtlety in quantum information is that our entropic measure must also accommodate the possibility of coherences between outcomes (non-diagonal elements in the density matrix).
Moreover, when extracting probabilities from a quantum state, there are many choices of measurement.
Thus, to be a function of a quantum state, rather than also of an (implicit) measurement,
 one typically defines a quantum entropy with respect to a minimization over all choices of projective measurement~\cite{NielsenC00}.
If one then calculates a R\'enyi entropy on these probabilities (or indeed, any concave function),
 the Schur-Horn theorem tells us that this minimization will be achieved
 when we measure in the diagonal basis of the density operator -- that is, when the outcome probabilities of the measurement are given by the eigenvalues of the quantum state.

For the case of the quantum generalization of the Shannon entropy, we can write this minimization directly in terms of $\rho$ to yield the {\em von Neumann entropy}
\begin{equation}
\label{eq:vonNeumann}
H_1\!\left(\rho\right) = -\tr \left(\rho\ln\rho\right).
\end{equation}
It is sometimes fashionable to introduce additional notation ($S$ and $H$) to distinguish the von Neumann entropy from the Shannon entropy.
We here view that as redundant mysticism,
 since classical probability theory is a subset of quantum information theory, and applying Eq.~\eqref{eq:vonNeumann} to the appropriate quantum representation $\rho_X$ of the classical random variable $X$ will yield exactly the same value as calculating the Shannon entropy directly.
Although some subtlety must be considered with mutual informations and conditional entropies,
 for unipartite systems the operational meaning of information entropy is the same between the classical and quantum cases:
 {\em Schumacher compression}~\cite{Schumacher95} quantizes Shannon's source coding theorem, demonstrating that $N$ independent copies of a quantum state $\rho$ can be transmitted in a Hilbert space (i.e.\ quantum configuration space) of size $N H_1\left(\rho\right)$, in the asymptotic limit $N\to\infty$.

Likewise, one can define~\cite{MuellerDSFT13} the family of quantum R\'enyi entropies associated with quantum state $\rho$: 
\begin{equation}
H_\alpha(\rho) := \dfrac{1}{1-\alpha} \ln \left[ \frac{ \tr{\rho^\alpha} }{\tr{\rho}} \right], \qquad \alpha \geq 0.
\end{equation}
When $\rho$ is chosen to encode a classical probability distribution, Eq.~\eqref{eq:RenyiEntropy} is recovered.
(The $\tr\rho$ in the denominator accommodates the possibility of unnormalized $\rho$.)

Similarly, there are quantum definitions for divergences.
For quantum states $\rho$, $\sigma$, the \KL{} divergence ($\alpha=1$) is defined as
\begin{equation}
\label{eq:QuantumKL}
\rDiv{1}{\rho}{\sigma} := \tr{\rho \left(\ln\rho -\ln\sigma\right)}.
\end{equation}

The general quantum R\'enyi relative entropy is expressed~\cite{MuellerDSFT13}:
\begin{equation}
\label{eq:QuantumDivergence}
\rDiv{\alpha}{\rho}{\sigma} := \dfrac{1}{\alpha-1}  \ln \left[ \frac{ \tr{\rho^\alpha \sigma^{1-\alpha}} }{\tr{\rho}} \right].
\end{equation}
Other quantum generalizations are possible, such as the {\em sandwiched quantum R\'enyi divergence}~\cite{MuellerDSFT13,WildeWY14,TomamichelBH14}, which coincides with the above when $\rho$ and $\sigma$ commute.
In this form, it can be straightforwardly seen~\cite{MuellerDSFT13} by setting $\sigma=\id$ that $H_\alpha{\left(\rho\right)} = -\rDiv{\alpha}{\rho}{\id}$---that is, the Renyi entropy itself can be considered as a divergence of $\rho$ from the (unnormalized) state $\id$.
Alternatively if we consider the uniform random state with the same dimension $d$ as $\rho$, $\pi := \id / d$,
 then we also have $H_{\alpha}\left(\rho\right) = \log d - \rDiv{\alpha}{\rho}{\pi}$.
Since $D_\alpha \geq 0$ (with equality holding for general $\alpha$ only when $\rho=\pi$), this also tells us that for all $\alpha$, the maximum value $H_\alpha$ can take is $\log d$ and that this is saturated when the quantum state is maximally mixed. That is, the uniformly random probability distribution maximizes all R\'enyi entropies for a given alphabet size.

We conclude this section by giving the explicit forms~\cite{TomamichelBH14} for the quantum R\'enyi relative entropies in the cases $\alpha=0$:
\begin{equation}
\label{eq:D0Quantum}
\rDiv{0}{\rho}{\sigma} = - \ln \tr \left( \pi_\rho \sigma \right)
\end{equation}
where $\pi_\rho$ is projection onto the support of $\rho$;
 and in the case of $\alpha\to\infty$:
\begin{equation}
\label{eq:DinftyQuantum}
\rDiv{\infty}{\rho}{\sigma}
= \ln \max_{i, j} \left\{ \dfrac{\lambda_i}{\mu_j} : \braket{i}{\tilde{j}}  \neq 0 \right\},
\end{equation}
where $\rho = \sum_i \lambda_i \ketbra{i}{i}$ and $\sigma = \sum_j \mu_j \ketbra{\tilde{j}}{\tilde{j}}$.
In all these cases, it is left as an exercise to the reader to show that when $\rho$ and $\sigma$ are classical probability distributions, these quantities reduce to their respective classical versions.

\section{Thermodynamic prerequisites}
\label{sec:ThermoTools}
\subsection{Information entropy in thermodynamics}
Having now equipped ourselves with some information-theoretic hammers, let us seek out some thermodynamic nails.
Broadly, there are two different contexts where information entropy appears in modern thermodynamics.
{\bf (i)} The entropy of the state of (small, possibly quantum) systems, undergoing some transformation. 
{\bf (ii)} The entropy of distributions of variables associated with thermodynamic processes, such as the total amount of work investment into a system as it undergoes some protocol.

These two contexts are conceptually very different. 
In the former, information entropy is akin to a thermodynamic state variable\footnote{A formal discussion on where the information entropy can be considered a thermodynamic entropy is presented by \citet{WeilenmannKFR16}.}, since it is something that can be assigned to a given system at a point in time.
The second context is more statistical in nature: the entropy here is not so associated with the system itself, but instead with data built up over many trials of an experiment.
In both these contexts, where Shannon entropies are used to quantify average properties, one may consider alternative statements involving R\'enyi entropies that allow statements to be made outside of this asymptotic limit.

In this article, we shall primarily focus on context (ii).
However, let us first make a few comments about (i), since it is a lively area of active research. 

Approach {(i)} follows the modern information thermodynamics paradigm established by \citet{Landauer61} and \citet{Bennett82,Bennett03} asserting the {\em inevitable physicality of information}~\cite{Landauer96}. 
This is particularly useful if we wish to make statements about the fundamental thermal costs of information processing.
Indeed, Maxwell's d\ae{}mon~\cite{Maxwell1870,LeffR02} -- an intelligent agent that can seemingly violate the second law of thermodynamics by converting heat to work without any other consequence -- is resolved by noting that this agent must maintain some kind of {\em memory},
 and this memory is itself a physical system that must be reset at some thermodynamic cost in order to have a complete thermodynamic cycle.

The exact lower bound on the cost of reconfiguring memory may be quantified by Landauer's principle~\cite{Landauer61},
 which relates the {\em Shannon entropy} of the memory -- an information--theoretical quantity --
 to the exchange of thermodynamic work -- a physical quantity, such as the raising or lowering of a weight.
Namely, consider a random variable $X$ corresponding to a choice of $x_i \in \mathcal{X}$, each occuring with probability $p_x(i)$,
 and a random variable $Y$ corresponding to $y_i \in \mathcal{Y}$ with probabilities $\{p_y(i)\}$.
Suppose some physical system $\Xi$ encodes the variable $X$.
If we wish to reconfigure $\Xi$ so that it encodes the variable $Y$, 
 if this reconfiguration is done in contact with a thermal reservoir at temperature $T$,
 then the {\em minimum} average work investment required is given by:
\begin{equation}
\label{eq:Landauer}
W \geq \kB T \left[ \Ent{X} - \Ent{Y} \right],
\end{equation}
where $\Ent{A}$ is the Shannon entropy of $A$.
If $\Ent{X} < \Ent{Y}$, then the right-hand-side is negative, signifying that work can be {\em extracted} from the physical system, at the cost of making it more random.
It follows from the first law that this work investment will be accompanied by an equal and opposite exchange of heat with the thermal reservoir.

As an example, consider {\em bit reset}: here, $X$ is a $\left(\frac{1}{2},\;\frac{1}{2}\right)$ distribution over two possibilities, whereas $Y$ reflects a pure state of certainty [probabilities $\left(1,\;0\right)$]. 
Thus $H(X) = \ln 2$, $H(Y) = 0$, and we recover the well-known ``Landauer erasure'' cost of $\kB T \ln 2$.

Various thought experiments, such as Szilard's engine~\cite{Szilard29,LeffR02} allow the derivation of this same cost from physical considerations (e.g.\ applying ideal gas laws to expanding pistons).
Should some previously unknown material be discovered, we would not expect that it could be used to form a perpetual motion machine: %
 the second law of thermodynamics likely holds independently of the underlying physical mechanisms.
Similarly, we can understand Landauer's principle as an {\em emergent physical law}
 that we expect to hold true no matter the specifics of the microscopic laws governing the system on which it is applied.
Systems that claim to violate this bound typically also draw from another type of reservoir that is freely convertible into work, such as information reservoirs, or move the entropy onto a different physical system (or into correlations) that is unaccounted for.
The work cost associated with a change of information entropy encoded on {\em any} physical medium will be subject to this limit\footnote{
Conversely, this limit can be sharpened by restricting the set of allowed protocols (such as by bounding the effective dimension of the thermal reservoir~\cite{ReebW14}).
}.

For any physically realisable experiment, a larger work investment than predicted from the Landauer bound will typically be required for two distinct reasons:
 {\bf (i)} Landauer's bound assumes a quasistatic change, where the physical system is kept in thermal equilibrium throughout the reconfiguration. 
This requires infinitely slow adjustments, whereas conversely, real experiments proceed in a finite amount of time.
 {\bf (ii)} The bound is derived on {\em asymptotic} information quantities: assuming either a large number of copies of the systems that can be compressed (i.e.\ such that the bound is a limit ``per system on average''), or many repetitions of the experiment where the statistics are collected and averaged, and the thermal reservoir is assumed to be infinitely large (whereas in fact it could also be of limited size~\cite{ReebW14}).

Non-equilibrium statistical mechanics can help us with reason (i), wherein the difference between entropies informs the free energy difference associated with states encoded in different distributions.
Conversely, {\em one-shot statistical mechanics} (sometimes called single-shot statistical mechanics) allows us to go beyond this average regime.
There are a wealth of papers on this topic (e.g.~\cite{DelRioARDV11,DahlstenRRV11,Dahlsten13,Aberg13,HorodeckiO13,EgloffDRV15,BrandaoHNOW15,WeilenmannKFR16}) that extend well beyond the scope of this chapter.
For a particularly pedagogical introduction, one could consult \citet{Dahlsten13}.

\subsection{Work extraction games}

\begin{figure}[hbt]
\begin{center}
\includegraphics[width=0.65\textwidth]{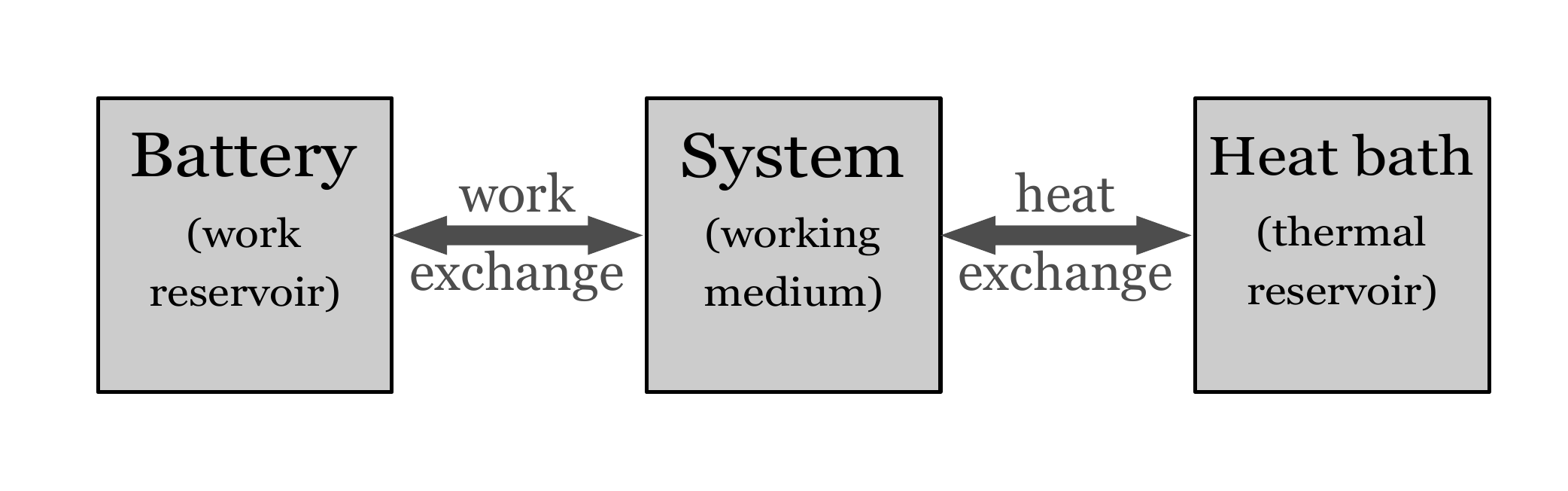}
\caption[Generic thermodynamic scheme]{
\textbf{Generic one-bath thermodynamic scheme.} 
A working medium (system) exchanges work with a battery (work reservoir), and heat with a heat bath (thermal reservoir).
}
\label{fig:Scheme}
\end{center}
\end{figure}

Let us now turn our attention to the second context in which entropies can appear in the discussion of thermodynamics.
We begin by setting out the following very general thermodynamic scheme (\cref{fig:Scheme}):
Suppose there is an experiment consisting of three components
 {\bf (a)} a working medium, or {\em system},
 {\bf (b)} a work reservoir, or {\em battery} (e.g.\ a raising weight),
 {\bf (c)} a heat reservoir, or {\em heat bath} at some temperature $T$.
On this set-up,
 one composes a {\em protocol}: a predefined sequence of actions taken on the system involving two types of interaction:
The first type is to allow some total-energy preserving interaction%
\footnote{
The reader should take care that there are varying notions of energy conservation, 
 and in resource-theoretic frameworks, this will alter the set of permitted ``thermal operations''. 
In particular, 
 one may admit any operation that conserves energy on average (as per \cite{SkrzypczykSP14}),
 or alternatively could place stricter restrictions, 
 such as mandating that the unitary representation of the dynamics on the system-battery commute with the total Hamiltonian (as per \cite{HorodeckiO13}).
Here, we present a general scheme that can be adapted to either notion.
} with the battery, 
 such that energy is exchanged between the battery and the system.
This energy change could either be due to some joint dynamic on the working medium and the battery (a favoured approach for truly quantum thermodynamics), or could be viewed abstractly as the imposition of a time-varying Hamiltonian on the system, where the battery acts as a generic work reservoir that can supply the resulting difference in internal energy on the system, such as by raising or lowering a weight\footnote{
Furthermore, these two pictures can be seen to be equivalent: a time-varying Hamiltonian can be recast as a time-invariant Hamiltonian with the help of an ancillary ``clock'' system -- for details, see Supplementary Material Section VIII of \citet{BrandaoHORS13},  Appendix D2 of \citet{YungerHalpernGDV15}, or Section IIA in \citet{AlhambraMOP16}.
}.
Energy exchanged in this manner is classified as {\em work}.

The second type of interaction is thermal contact between the system and the heat reservoir (though the system need not necessarily fully equilibrate).
Energetic exchanges resulting from this are considered {\em heat}.

In an experiment, one can perform many {\em trials} of this protocol, and each trial will exchange some amount of work $W$ with the battery, and some amount of heat $Q$ with the heat bath.
From the collated statistics of $W$ over many trials, one can form a {\em work distribution} $P(W)$.
(One might also infer $P(W)$ from the theoretical details of a given protocol).
 
This scheme is sufficiently general that it describes the thermodynamic behaviour of any system interacting thermally with a single heat bath and drawing upon (or depositing into) a work reservoir.
We can flesh it out with some details to specialize to the microscopic quantum case where the working medium has discrete energy levels (following the lead of, say, \cite{Aberg13,Dahlsten13,AndersG13}).
Here, the working medium is characterised by 
 its time-varying Hamiltonian $H(t) := \sum_i E_i(t) \ketbra{E_i(t)}{E_i(t)}$,
 and its {\em state} $\rho(t)$, which represents a distribution over energy levels.
If the system is classical, then $\rho = \sum_i p_i(t) \ketbra{E_i(t)}{E_i(t)}$ is diagonal with respect to the Hamiltonian basis.

The system's average internal energy is given $\expt{H} = \tr{\rho H} = \sum_i p_i E_i$.
If the system is classical, its state remains diagonal with respect to the Hamiltonian at each timestep (e.g.~by actively avoiding the build up of quantum coherence~\cite{BrowneGDV14}), and hence may always be expressed as a classical distribution over energy levels.
In this case, the total differential change to the energy may be expressed
\begin{equation}
d\!\expt{H} = \sum_i p_i dE_i + \sum_i E_i dp_i.
\end{equation}
By analogy with the first law $dU = dW + dQ$ we can divide these two terms into~\cite{Alicki79,Piechocinska00,AlickiHHH04,Aberg13}:
\begin{enumerate}
\item {\bf Exchanges of work}: $dW = \sum_i p_i dE_i$, associated with changing the Hamiltonian at fixed occupation probability.
This is motivated as work, since it corresponds to a change in energy as a consequence of some external action on the system~\cite{Schrodinger46} (i.e.\ arises due to the time-dependence of the Hamiltonian) such as adjusting an external magnetic field or raising a weight.
\item {\bf Exchanges of heat}: $dQ = \sum_i E_i dp_i$, associated with changing the populations at fixed Hamiltonian.
This is motivated as heat, since it corresponds to the changes in energy effected by thermalization as a system is moved towards its equilibrium state~(see e.g.\ \cite{HorodeckiO13,YungerHalpernGDV15}).
\end{enumerate}
In general, a protocol is constructed by inducing a combination of the above two classes of operations.
Physical considerations may lead us to impose further constraints, particularly to ensure that the allowed heat-like exchanges obey the second law.
An obviously motivated restriction (see, e.g.\ \cite{HorodeckiO13}) is that heat-like changes should move the system towards a thermal equilibrium state (the {\em Gibbs state} $\gamma = \diag{\left[ e^{-\beta E_1}/Z, \ldots, e^{-\beta E_N}/Z \right]}$ where $Z := \sum_i e^{-\beta E_i}$ is the partition function.)
We can further restrict heat-like interactions to obey {\em detailed balance}\footnote{
This constraint is closely related to first, see Appendix A of \cite{YungerHalpernGDV15}.
} such that the ratio of transition probabilities between two configurations $i$ and $j$ obeys $\sfrac{P(i\to j)}{P(j\to i)} = \exp{\left[-\beta\left(E_j - E_i\right)\right]}$.

\begin{figure}[bht]
\begin{center}
\includegraphics[width=0.85\textwidth]{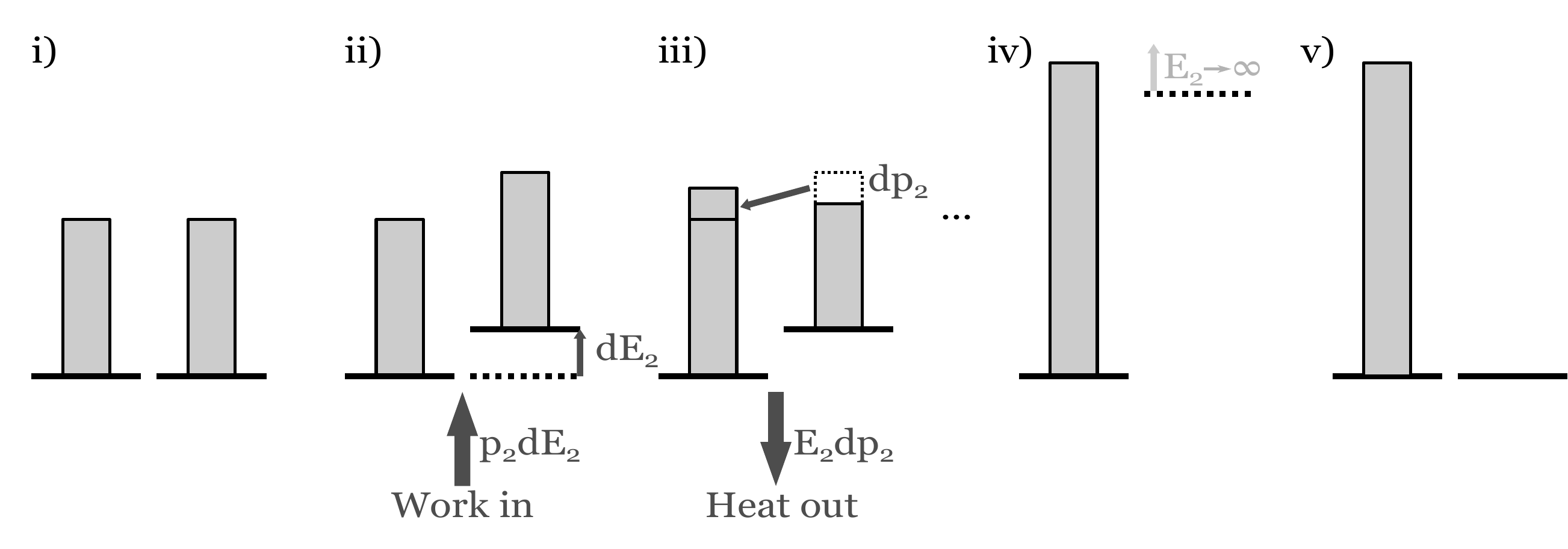}
\caption[Landauer bit reset]{
\textbf{Landauer bit reset.} 
A diagrammatic representation of the bit reset protocol (following notation of \citet{Aberg13}), representing the system's state and Hamiltonian throughout the protocol.
The two thick horizontal lines represent energy levels, with their relative vertical position representing the energy of each level. The gray bars above each level represents its occupation probability.
i) Initially the system is in an equal mix of degenerate energy levels.
ii) At each time-step, the second level is raised by $dE_2$, at work cost $p_2 dE_2$.
iii) The system is then left to thermalize such that the probability of the second level being occupied changes by $dp_2$. This exchanges heat $E_2 dP_2$ with the heat reservoir.
iv) Steps ii and iii are repeated until the second level is raised to infinity, and hence totally depopulated.
v) The system is then decoupled from the thermal reservoir, such that the (empty) second level can be set back to $0$ energy without any associated energy change.
The total work cost over the entire protocol is $\kB T \ln 2$.
}
\label{fig:BitReset}
\end{center}
\end{figure}

As an example, let us consider quasistatic {\em bit reset}: the process of taking a two-level system from state $\rho = \diag\left(\frac{1}{2},\;\frac{1}{2}\right)$ to $\rho=\diag\left(1,\; 0\right)$, where the initial and final Hamiltonian is $H=0$.
This is expressed diagrammatically in \cref{fig:BitReset}.
The theoretically-perfect quasistatic protocol for reset is to slowly increase the second level to $\infty$, allowing the system to fully equilibrate (that is, to reach the Gibbs state $\gamma = \diag \left[\sfrac{1}{\left(1+e^{-\beta E_2}\right)},\;\sfrac{e^{-\beta E_2}}{\left(1+e^{-\beta E_2}\right)} \right]$, where $\beta =\sfrac{1}{\kB T}$ is the inverse temperature) between each infinitesimal change in $E_2$. 
Once the second level has been fully raised, the system is decoupled from the thermal reservoir, and the (now empty) second level is lowered back to $0$, completing the protocol.
The latter stage has no associated work cost, since it is a change in energy of a completely depopulated level, and so the total expected work cost $\expt{W_{\rm reset}}$ of the protocol is given by the first stage:
\begin{equation}
\label{eq:LandauerBitReset}
\expt{W_{\rm reset}} = \int_0^\infty p_2 dE_2 = \int_0^\infty\dfrac{e^{-\beta E_2}}{1+e^{-\beta E_2}} dE_2 = \frac{1}{\beta}\ln 2 = \kB T \ln 2.
\end{equation}
This work cost exactly matches that predicted by applying Landauer's principle on the change between initial and final entropies [Eq.~\eqref{eq:Landauer}], indicating that this protocol is optimal.
A similar integration over population changes will yield $\expt{Q_{\rm reset}} = -\kB T \ln 2$, just as we would expect from the first law (since $\Delta U = 0$).
One can also formulate variations of this protocol outside of the quasistatic limit, by accommodating the possibility that the system does not thermalize completely between each change in the energy of the second level~\cite{BrowneGDV14}.

These quantities correspond to average behaviours.
When the system is classical, 
 it is straight-forward to generalize this to a {\em trajectory formalism} (as per \citet{Crooks98} or in \cite{Aberg13,Dahlsten13}),
 and talk about single trials of the experiment.
Here, the system is taken to occupy one particular energy level at any given time (i.e.\ so that the density matrix is always a rank one projector in the energy eigenbasis).
In this picture, thermalization corresponds to random jumps between energy levels.
A single trajectory is then a ``path'' through the system's state space (i.e.\ a list of energy levels occupied and the times of transitions between them).
Work and heat costs may then be calculated for a given trajectory,
 such that the mean values correspond to the average over all trajectories.

\subsection{Worst-case \mbox{and $\epsilon$-guaranteed work}}
When we use heat and work in the context of traditional thermodynamic processes,
 we typically refer to the mean (``average'' or ``expected'') work $\expt{W}$ (cf.\ heat $\expt{Q}$) taken over many runs of an experiment.
If the thermodynamic protocol acts on a large system and is quasistatic (evolving sufficiently slowly that the system remains in internal equilibrium),
 the work costs of individual runs of the experiment do not deviate much from the mean values.
However, when the protocol happens over a finite amount of time, the system can be driven out of thermal equilibrium, 
 and the work cost of an individual run of the experiment can deviate quite significantly from the mean value.
Here, one instead records a {\em work distribution} $P(W)$---a probability density function over the work costs of each run of the experiment.
From this distribution, one can calculate the mean work $\expt{W}$ in the usual way: $\expt{W} = \int_{-\infty}^{\infty} dW\;W\,P(W)$.

There are circumstances where mean properties of a distribution are misleading.
Consider a game where a ball is launched upwards from the ground, with the aim that it reaches a table of height $1\,{\rm m}$.
Suppose with probability $0.99$ the ball reaches only reaches $0.1\,{\rm m}$, but with probability $0.01$ it is launched to $100\,{\rm m}$.
The mean height reached by this ball is $1.099\,{\rm m}$, which is greater than the table height -- but the ball will not reach the table in $99\%$ of launches.
As such, the mean is not a good characterisation of the process for the purposes of determining whether some threshold is met.

Similar circumstances arise in a thermal contexts with respect to work distributions $P(W)$.
It could be that a single instance of the experiment is powered by a battery that can only impart so much work in one run, 
 or the system contains some small wire (such as might be found in the filaments of microprocessor's transistors) that can only support up to a maximum amount of heat dissipation before it overheats and breaks.
In these contexts, we are less concerned by the average work $\expt{W}$, but rather wish to know the maximum work $W_{\rm max} := \max_W \{W | P(W)>0\}$.
In protocols whose net work exchange is to invest work from the battery into the system,
 this maximum quantity is sometimes referred to as the {\em worst--case work cost}.

Conversely, for processes where work is typically extracted out of the system and into the battery, the ``maximum'' quantity will be the least negative.
Here, the process could be providing the input energy to trigger another process with a minimum activation energy: 
 an ``all or nothing'' threshold that must be crossed in order to start a chemical reaction (similar to the ball and table example above).
In this extraction context, $-W_{\rm max}$ is sometimes known as the {\em guaranteed work}, 
 since each trial of the protocol is guaranteed to output at least that much work. 
(Here, extracting more work is desirable, so the term ``worst--case'' would still be an appropriate adjective).

\begin{figure}[htb]
\begin{center}
\includegraphics[width=0.45\textwidth]{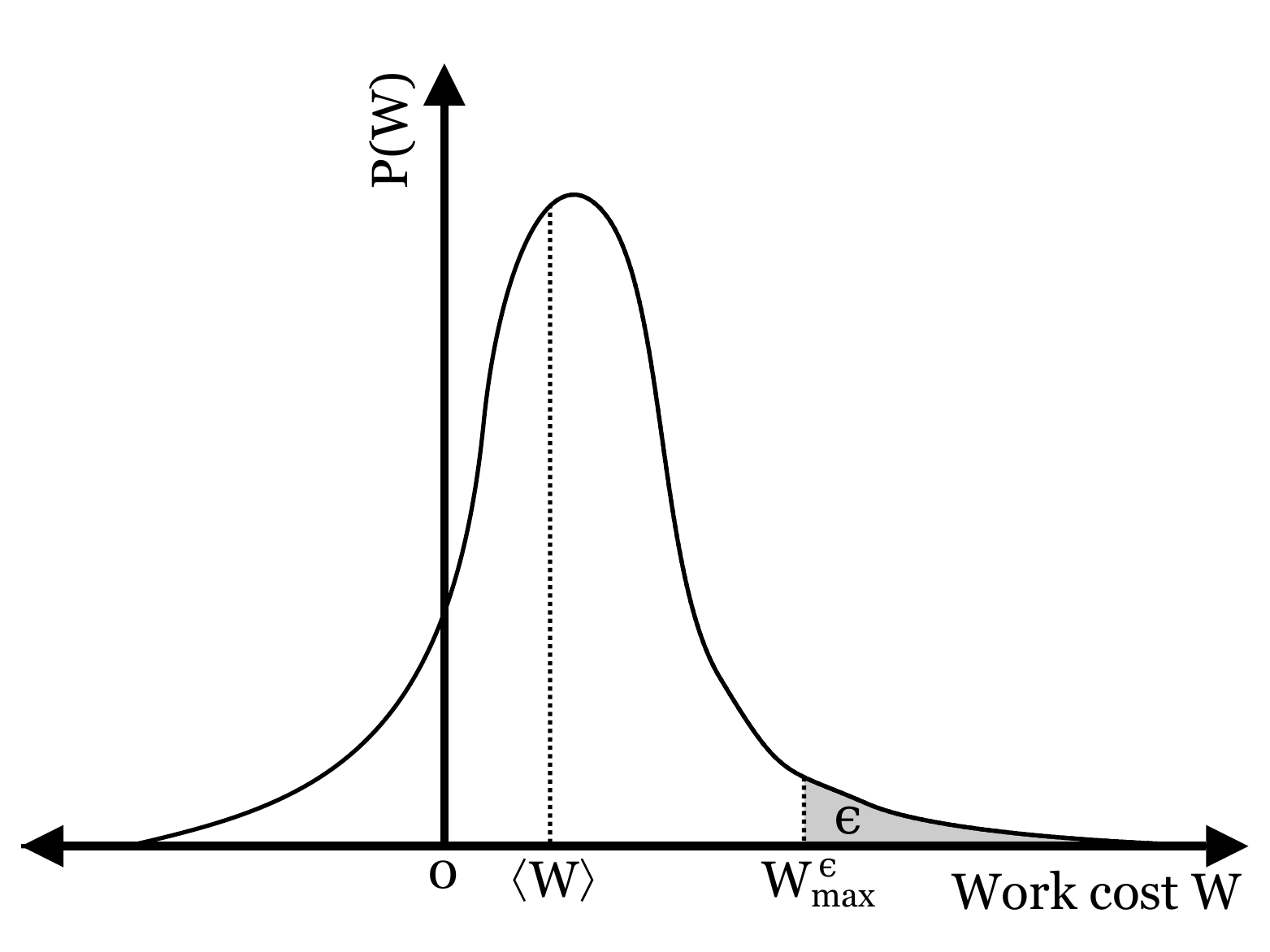}
\caption[Worst-case work]{
\textbf{Example work distribution,}
 with the mean work $\expt{W}$ and the the $\epsilon$-guaranteed worst-case work $W_{\rm max}^\epsilon$ marked.
Positive $W$ indicates work must be invested from the battery into the system.
The grey area shaded region has area $\epsilon$, such that the probability that a work sampled from this distribution exceeds $W_{\rm max}^\epsilon$ is $\epsilon$.
}
\label{fig:WorstCaseWork}
\end{center}
\end{figure}

Sometimes, we are not interested in the {\em absolute} worst-case scenario, but can tolerate some {\em failure probability} $\epsilon$.
The {\em $\epsilon$-guaranteed worst--case work cost} is the lowest value $W_{\rm max}^{\epsilon}$ that satisfies
\begin{equation}
\int_{W_{\rm max}^\epsilon}^\infty \hspace{-1em} dW \; P(W) = \epsilon.
\end{equation}
(This is drawn in \cref{fig:WorstCaseWork}).

One expects the work cost of a trial to exceed $W_{\rm max}^{\epsilon}$ only with probability $\epsilon$.
This is a generalization of the work distribution's median value; setting $\epsilon = \frac{1}{2}$ yields the median work.
This definition also holds for processes where work is output (though we must pay attention to the signs).
Should we alternatively express such a process in terms of $-W$ (i.e.\ exchanging the sign),
 we have
\begin{equation}
\int_{-\infty}^{W_{\rm worst}^\epsilon} \hspace{-1em} dW \; P(-W) = \epsilon,
\end{equation}
where $W_{\rm worst}^\epsilon$ is the {\em $\epsilon-$guaranteed work output}.
Precisely what {\em failure} means is context-dependent: it could mean that in $\epsilon$ of cases the protocol completes, but costs more work than allowed; or it could mean that in $\epsilon$ of cases a different protocol is executed (say, because the battery is drained).

In the case of quasistatic processes on large systems (particularly, on ensembles of many systems that are independent or otherwise have correlations that are limited in range),
 the work distribution $P(W)$ will be (close to) a Dirac delta function $P(W) = \delta(W-\Delta F)$ (where $\Delta F$ is the free energy change).
Here, $\expt{W}\approx W_{\rm max}^\epsilon$ for all $\epsilon$, and knowing the mean work is sufficient to characterize the protocol's work cost.
In the next section we shall discuss the far more interesting context wherein the system is driven out of equilibrium,
 and the work distribution is no longer well-characterized by the mean.

\section{Work fluctuation theorems and one-shot entropies}
\label{sec:WorkFluctuations}
Fluctuation theorems are powerful tools that allow us to make statements relating out of equilibrium behaviour with equilibrium properties such as free energy differences.
Of particular interest is Crooks' fluctuation theorem~\cite{Crooks99},
 that relates the entropy production of a system driven in one direction, with that of the reversely driven system.
This relation holds under the assumption that each individual microscopic trajectory the system takes is {\em reversible}.
In this context, {\em microscopic reversibility} means that for each forward process trajectory there exists a corresponding reverse trajectory in the reversed process that moves through the same states but in time-reversed order, 
 and that the ratio of probabilities that the system evolves according to these trajectories in their respective process directions is given by $\exp\left(-\beta Q\right)$ where $Q$ is the heat exchanged with the thermal reservoir during the forward trajectory~\cite{Crooks99}.
Microscopic reversibility is satisfied if the driving of the protocol is characterized by a single parameter, and the thermalizing interactions obey detailed balance~\cite{Crooks98}.

The killer application of Crooks' theorem (provided also in \cite{Crooks99}) is to model the probabilistic work cost of out-of-equilibrium thermodynamic processes.
Suppose there is a process that begins in a thermal state (with respect to initial Hamiltonian), and undergoes evolution driven by a single external parameter $\lambda(t)$ (over range $0$ to $\tau$), and where all thermalizing  interactions obey {\em detailed balance}.
Such a process has a well-defined {\em reverse},  wherein the driving parameter $\lambda$ is varied from $\tau$ back to $0$, 
 and the initial state of the reverse is the thermal state of the system where the external parameter is $\tau$ (this is not necessarily the same as the final state of the forward process, since in general the system may have been moved out of thermal equilibrium).
When the evolution is not quasistatically slow, 
 there is an element of randomness to the {\em work cost} $W$ of these processes. 
Describing the distribution over work cost in the forward case by $P_{\rm fwd}$ and in the reverse as $P_{\rm rev}$, 
 Crooks' fluctuation theorem~\cite{Crooks99} implies\footnote{
Sometimes Eq.~\eqref{eq:Crooks} is referred to as Crooks' theorem itself. However, in \cite{Crooks99}, Crooks proves a more general statement about entropy production systems with microscopically reversible dynamics: $P_{\rm fwd}\left(\omega\right)/P_{\rm rev}\left(-\omega \right) = \exp\left(\omega\right)$. The equation pertaining to work exchanges is the most notable example, and was also provided by Crooks' in the same article.
} the {\em nonequilibrium work relation}:
\begin{equation}
\label{eq:Crooks}
\dfrac{P_{\rm fwd}(W)}{P_{\rm rev}(-W)} = \exp\left[\beta\left(W - \Delta F\right)\right],
\end{equation}
where $\Delta F$ is the equilibrium free-energy change\footnote{
Under the assumptions that allow us to use Crooks' nonequilibrium work relation, the initial state is in thermal equilibrium and its free energy may be defined in the usual way.
At time $\tau$, the system may not be in thermal equilibrium. 
However, we can consider the state that would be reached (for the same final Hamiltonian) if that system were allowed to fully thermalize (limit $t\to\infty$), and take the free energy of that thermal state instead. 
It is the difference in free energy between the initial thermal state and the thermalized version of the final state that determines $\Delta F$. 
} associated with the forward process.

Although we here shall not supply a full proof of the equality\footnote{
\citet{QuanD08} provide a derivation of this in a language familiar to quantum information scientists.
}, we can see intuitively why the free energy difference appears:
 our set-up involves two thermal states, and these effectively encode within them the partition function $Z$ associated with their respective Hamiltonians.
Elementary statistical mechanics tells us that $F = -\kB T \ln Z$, and so a logarithm of the ratio of partition functions (as will be generated when we sum over the probabilities associated with the two thermal states) will supply terms of the form $\Delta F$.

Crooks' fluctuation theorem implies other important thermodynamic statements.
For instance, multiplying Eq.~\ref{eq:Crooks} by $e^{\beta \Delta F} P_{\rm rev}(-W)$ then integrating over $W$ yields the famous Jarzynski equality~\cite{Jarzynski97}:
\begin{equation}
\label{eq:Jarzynski}
\expt{e^{\beta W}} = e^{\beta \Delta F}.
\end{equation}
In turn, applying Jensen's inequality to Eq.~\eqref{eq:Jarzynski} yields a statement of the second law in the form $\expt{W} \geq \Delta F$.

The energetic quantity appearing on the right hand side of Eq.~\eqref{eq:Crooks} is known as the {\em dissipated work} $W_{\rm diss} := W - \Delta F$.
The {\em dissipated work} quantifies the excess work investment required to complete a protocol beyond the free energy difference between the initial and final states. 
Since $\Delta F$ is a property of the protocol (rather than of trials), it is the same for every run of an experiment, 
 and hence one can define the {\em average dissipated work} $\expt{W_{\rm diss}} = \expt{W} - \Delta F$.

This dissipated work represents a wastefulness caused by taking the system out of thermodynamic equilibrium.
The free energy difference of a process and its reverse are related by simple negation.
However, $\expt{W_{\rm diss}}\geq0$ in {\em both} directions.
(If this were not true, one could close either of these processes into a cycle with the dissipation-free quasistatic variant of the reserve, and violate the second law.)
This follows also from Crooks' fluctuation theorem~\cite{WuK05,GomezPB08}.
Recall the definition of the \KL{} divergence (Eq.~\eqref{eq:crd1}), and substitute in Eq.~\eqref{eq:Crooks}:
\begin{align}
\rDiv{1}{P_{\rm fwd}(W)}{P_{\rm rev}(-W)} & = \int_{-\infty}^{\infty} \hspace{-0.5em} dx \; P_{\rm fwd}(W) \ln \frac{P_{\rm fwd}(W)}{P_{\rm rev}(-W)} \nonumber \\
& = \int_{-\infty}^{\infty} \hspace{-0.5em} dx \; P_{\rm fwd}(W) \beta \left[W - \Delta{F}\right] \nonumber \\
& = \beta \expt{W_{\rm diss}}.
\end{align}
Thus, the \KL{} divergence between the distribution of work invested into the forward process, and the distribution of work extracted from the reversed process is directly proportional to the average dissipated work.
$\expt{W_{\rm diss}}\geq0$ then follows from the positivity of $D_1$.
Moreover we see that $\expt{W_{\rm diss}}=0$ if and only if $P_{\rm fwd}(W) = P_{\rm rev}(-W)$ for all $W$,
 which provides us with a slightly more generic definition of thermodynamic reversibility~\cite{GomezPB08}.

\citet{YungerHalpernGDV15_WC} show that a similar relation holds true for the {\em worst-case dissipated work}, defined $W_{\rm diss}^{\rm worst} := W_{\rm max} - \Delta F$:
\begin{equation}
W_{\rm diss}^{\rm worst} = \kB T \rDiv{\infty}{ P_{\rm fwd}(W) }{ P_{\rm rev}(-W) }.
\end{equation}
Similarly to $D_1$, one proves this by substituting Crooks' nonequilibrium work relation into the definition of $D_\infty$.
The protocol's adherence to Eq.~\eqref{eq:Crooks} guarantees that distributions $P_{\rm fwd}(W)$ and $P_{\rm rev}(-W)$ have the same support.
Therefore, the largest value of $W$ in the forward process $W_{\rm max}$ is exactly the one which maximizes the right-hand side of Eq.~\eqref{eq:Crooks},
 and hence determines the minimum value of $\lambda$ in the right-hand-side of Eq.~\eqref{eq:crdInfty}.

A related equality is derived by \citet{DahlstenCBGYV17}.
Consider a system in an initial state $\rho_0$, which does not need to be a thermal state of the initial Hamiltonian, but must be diagonal in that basis.
Let that system evolve by some process $\mathcal{P}_{\rm fwd}$, consisting of changes in energy levels, and thermalizations that respect detailed balance. 
Ultimately, one arrives at a bound of the form $W < W_{\rm max}^{\varepsilon}$ where
\begin{equation}
\label{eq:OscarWork}
W_{\rm max}^{\varepsilon} = \kB T \left[ \rDiv{\infty}{\tilde{P}_{\rm fwd}(W)}{\tilde{P}_{\rm rev}(-W)} - \ln \left( \frac{Z_f}{\tilde{Z}}\right)\right],
\end{equation}
consisting of two terms, which we shall explain.
The first term is the worst case dissipated work encoded between the forward and reverse work distributions (as in \cite{YungerHalpernGDV15_WC}, discussed above).
The tilde above $\tilde{P}$ indicates a smoothed work distribution (see \cite{DahlstenCBGYV17} for detail) formed from the true work distribution by discounting trajectories in the following ``tails'':
 {\bf (i)} the values which are unlikely to be seen since they have very low support in the initial state $\rho_0$ (total probability $p_{\rm out}$);
 {\bf (ii)} the highest (i.e.\ worst-case) work values beyond some tunable cut-off value.
This is a type of {\em smoothing} and is done to lessen the influence of very unlikely trajectories,
 and to explicitly incorporate a degree of error we are willing to tolerate in the bound.

The second term is a modification of the usual free energy difference.
$Z_f$ is the partition function of the final Hamiltonian, 
 but $\tilde{Z}$ is some modified function that replaces the role of the initial partition function $Z_i$.
Particularly, $\tilde{Z}$ omits terms corresponding to energy levels that are in the set {\bf OUT} of system configurations that have the least support in the initial state $\rho_0$.
That is $\tilde{Z} = \sfrac{1}{(1-p_{\rm out})}\sum_{i \notin {\bf OUT}} \exp{\left(-\beta E_i\right)}$,
 where $p_{\rm out}$ is the total support of $\rho_0$ on these unlikely {\bf OUT} states.
Using $\tilde{Z}$ instead of the true initial partion function will provide us a bound on the worst case work, since one can then construct a process $\tilde{\mathcal{P}}$ whose work costs upper bound that in the true process $\mathcal{P}$, but that begins in the thermal state $\tilde{\gamma}$ associated with $\tilde{Z}$ (details in \cite{DahlstenCBGYV17}).

In this context, it is unimportant that the initial state be thermal,
 because we concerned only with a single trajectory (namely, the non-excluded trajectory with the worst work cost)  without regards to how likely it is to occur,
 and the support of $\rho_0^{\bf IN}$ ($\rho_0$ omitting terms in {\bf OUT} and renormalized) and that of the thermal state of $\tilde{Z}$ are identical.

For Eq.~\eqref{eq:OscarWork} to be useful, we must also bound the failure probability $\varepsilon$.
Let $p_{\rm work-tail}$ be the probability that a trajectory is culled for being too expensive, and recall that the probability that the system is in {\bf OUT} is $p_{\rm out}$.
We then see the total error is bounded $\varepsilon \leq p_{\rm work-tail} + p_{\rm out}$ (and will be strictly less if some of the worst-case work trajectories begin in an {\bf OUT} state).
However, $p_{\rm work-tail}$ is not quite the same as the $\epsilon$ guarantee discussed earlier, since the former quantity relates to the smoothed $P(W)$ arising from modified process $\tilde{\mathcal{P}}$, whereas the $\epsilon$ is defined with respect to the true work distribution $\mathcal{P}$.
However, we can bound $p_{\rm work-tail} \leq d\left(\rho_0, \tilde{\gamma}\right) + \epsilon$ where $d\left(\cdot, \cdot\right)$ is the trace distance, and $\epsilon$ the ``guarantee''.
As such, the total error probability is upper bounded by:
\begin{equation}
\varepsilon \leq p_{\rm out} + d\left(\rho_0, \tilde{\gamma}\right) + \epsilon.
\end{equation}

In another approach from \citet{YungerHalpernGDV15},
 a one-shot entropy is combined with Crooks' nonequilibrium work relation to bound the worst-case work of a process with respect to an entropic property of the reverse process's work distribution.
Particularly, suppose there is some process $\mathcal{P}_{\rm fwd}$ with a distribution over work costs given $P_{\rm fwd}(W)$.
We can bound the $\epsilon-$guaranteed worst-case (i.e.\ lowest) work $W^{\epsilon}_{\rm worst}$ that can be extracted from the reverse process $\mathcal{P}_{\rm rev}$  by
\begin{equation}
\label{eq:WorstCaseBound}
W^{\epsilon}_{\rm worst} \leq \Delta F - \kB T \left[ H^{\kB T}_\infty (P_{\rm fwd}) + \log\left(1 + \epsilon\right) \right].
\end{equation}
where $H_{\infty}^{\kB T}$ is the differential min-entropy (order-$\infty$ R\'enyi entropy) taken with respect to a reference probability distribution with width $\kB T$ and height $\beta$ (see discussion at the end of \cref{sec:DiffEntropy}). %
The proof follows by first deriving an minor modification to Jarzynski's equality
\begin{equation}
\left(1-\epsilon\right) e^{-\beta \Delta F} = \int_{W^{\epsilon}_{\rm worst}}^\infty \hspace{-1em} P_{\rm fwd}(W) e^{-\beta W}
\end{equation}
by rearranging Eq.~\eqref{eq:Crooks}, and integrating over $W_{\rm worst}^{\epsilon}$ to $\infty$ (cf.\ the lower limit $-\infty$, which would recover the standard Jarzysnki equality).
One can then bound the integral on the right-hand side using the largest value of $P_{\rm fwd}(W)$, $P_{\rm fwd}^{\rm max}$:
\begin{align}
\int_{W^{\epsilon}_{\rm worst}}^\infty \hspace{-1em} P_{\rm fwd}(W) e^{-\beta W}
& \leq P_{\rm fwd}^{\rm max} \int_{W^{\epsilon}_{\rm worst}}^\infty \hspace{-1em} dW e^{-\beta W} \nonumber \\
& = \exp\left(\ln \left( P_{\rm fwd}^{\rm max} / \beta\right) \right) \exp\left(-\beta  W_{\rm worst}^{\epsilon} \right) \nonumber \\
& = \exp\left( -H_\infty^{\beta}\left(P_{\rm fwd}\right) -\beta  W_{\rm worst}^{\epsilon} \right),
\end{align}
and then rearrange to arrive at Eq.~\eqref{eq:WorstCaseBound}.

\citet{YungerHalpernJ16} apply Ineq.~\eqref{eq:WorstCaseBound} to upper bound the estimate on the number of trials required to estimate the free energy difference of a process in terms of the min-entropy $H_{\infty}^{\kB T}\left[P(W)\right]$.
From Jarzynski's equality~\cite{Jarzynski97}, it had been established~\cite{Jarzynski06} that around $N \sim \exp\left(\beta \expt{W_{\rm diss}}\right)$ experimental runs are required to form a good estimate of $\Delta F$.
In particular, averages of the exponential quantity $\expt{e^{\beta W}}$ are dominated by large values of $W$.
Thus, to get a good estimate, one hopes to sample one of these large values.
Since these large values appear in the tail of the distribution, they may not be likely, and so one might need to take many samples before one appears: namely, requiring a number of trials proportional to the inverse of the tail's total probability.

Consider sampling the work cost of some reverse $\mathcal{P}_{\rm rev}$ of some process $\mathcal{P}_{\rm fwd}$.
One may wish to pick some $\delta-$guaranteed output work value $W^\delta$ (known in this context as a $\delta$-dominant work value), wherein the probability of an experimental trial outputting less work than that is $\delta$.
Then, approximately $N_{\delta} \sim \frac{1}{\delta}$ trials will be performed before such a trial is encountered.
One can rearrange Eq.~\eqref{eq:WorstCaseBound} and lower bound this number:
\begin{equation}
\label{eq:NicoleBound}
N_{\delta} \geq \exp\left[\beta\left(W^\delta - \Delta F\right) + H_{\infty}^{\kB T}\left(P_{\rm fwd}\right)\right].
\end{equation}
If $W^{\delta}$ is chosen such that $W^{\delta} = \expt{W}_{\rm fwd}$, 
 then 
$N_{\delta} \geq \exp\left[\beta\left(\expt{W_{\rm diss}}\right) + H_{\infty}^{\kB T}\left(P_{\rm fwd}\right)\right]$.
Furthermore, if the probability density $P_{\rm fwd}(W)$ is less than $\beta$, then $H_{\infty}^{\kB T}\left(P_{\rm fwd}\right)\geq 0$,
 and the bound on $N_{\delta}$ from Eq.~\eqref{eq:NicoleBound} will be tighter than that implied by \citet{Jarzynski06}.

\newpage
\section{Quantum fluctuation theorems and one-shot entropies}
\label{sec:QuantumFluctuations}
Let us conclude this chapter by considering a few instances in which one-shot entropies have been applied to {\em quantum fluctuation relations}. 
It was shown by \citet{Kurchan00} and \citet{Tasaki00} that Jarzynski's equality can be applied in the quantum realm for a restricted set of protocols
 where a quantum system begins in a thermal state with respect to its initial Hamiltonian,
 and undergoes unitary evolution governed by a time-dependent Hamiltonian $H(t) = \sum_i E_i(t) \ketbra{E_i(t)}{E_i(t)}$ from time $t=0$ to $\tau$.
At the end of this evolution, the system is projectively measured in the basis of the final Hamiltonian $\tilde{H} := H(\tau)$. 
In these protocols, since the system is decoupled from the thermal reservoir throughout its evolution,
 one may call the energy difference between projective energy measurements made in $H(0)$ at the beginning and $H(\tau)$ end of the protocol the work cost of the protocol.
In this setting, if the system begins in energy level $i$ and ends in energy level $j$, the work cost would be $W = E_j(\tau) - E_i(0)$.
Following these rules, one ends up with the Jarzynski equality wherein $\Delta F$ corresponds to the difference in the free energy between the initial (thermal) state, and the thermal state associated with the final Hamiltonian $H(\tau)$.

An especially brief proof that such a set-up will obey the Jarzynski equality is supplied by \citet{Vedral12}, who points out that the equality holds essentially because quantum transformations conserve total probability.
Consider a system prepared initially in state $\rho_0 = \sum_i P(i) \ketbra{i}{i}$, and subjected to evolution by the unitary $U$.
After this evolution, the system is measured in the basis of projective measurement $\{\ket{j}\}$.
According to the Born rule, the joint probability of starting in $\ket{i}$ and ending in $\ket{j}$ is given
\begin{equation}
P(i,j) = \tr \left[ \ketbra{{j}}{{j}} U  \ketbra{i}{i} \rho_0 \ketbra{i}{i} U^\dag  \ketbra{{j}}{{j}} \right]
 = \tr \left[ \ketbra{{j}}{{j}} U \ketbra{i}{i} U^\dag \right] \times \tr\left[\ketbra{i}{i}\rho_0\right].
\end{equation}
The factorisation on the right-hand side expresses a simple probabilistic chain rule,
\begin{equation}
P(i,j) = Q(j|i) P(i),
\end{equation}
where $P(i)$ is the probability of beginning in state $i$, and $Q(j|i)$ is the {\em conditional probability} of ending in state $j$ given we started in $i$.
The final probability of interest, $Q(j) = \tr \left[ \ketbra{{j}}{{j}} U \rho_0 U^\dag \right]$, is the probability that the system is measured in state $j$, given that it was initially prepared according to the density matrix $\rho_0$.
One can define a (simple, classical) mutual information between the outcomes of the two measurements 
\begin{equation}
I_{ij} = -\ln Q(j) + \ln Q(j|i),
\end{equation}
 and then calculate
\begin{equation}
\expt{e^{I_{ij}}} = \sum_{ij} P(i,j) \exp\left({I_{ij}}\right) = \sum_{ij} P(i,j) \frac{Q(j)}{Q(j|i)} = \sum_{ij} P(i) \times Q(j) = \sum_i P(i) \times \sum_j Q(j) = 1.
\end{equation}

Suppose now that one takes the initial measurement as the Hamiltonian $H\!\left(0\right)$, the initial state as the thermal state of that Hamiltonian, and the final measurement as the Hamiltonian $H\!\left(\tau\right)$.
After some minor algebra, we find that $I_{ij} = \beta\left[E_i\left(0\right) - E_j\left(\tau\right)\right] - \beta \Delta F$, and since $\expt{e^{-I_{ij}}} = 1$, this recovers Jarzynski's equality for quantum systems.

Such a quantum process has a naturally defined {\em reverse process}, wherein $\tilde{H}(t) := H(\tau - t)$.
The reverse process state is written as $\tilde{\rho}(t)$, and the initial state $\tilde{\rho}(0)$ is here the thermal state of $\tilde{H}(0)$ (i.e.\ the associated thermal state with the final Hamiltonian $H(\tau)$ of the forward process).
\citet{ParrondoBK09} demonstrate that for such a process and its reverse, a quantum fluctuation theorem can be used to determine the mean dissipated work:
\begin{equation}
\label{eq:DissQWork}
\expt{W_{\rm diss}} = \kB T \rDiv{1}{\rho(t)}{\tilde{\rho}(\tau- t)}.
\end{equation}
The left-hand side has no time dependence (this would not make sense, since $\expt{W_{\rm diss}}$ is a property of the whole protocol rather than of one instance in time).
Indeed, the choice of time in $\rho$ / $\tilde{\rho}$ does not matter: $\rDiv{1}{\rho(t)}{\tilde{\rho}(\tau-t)}$ is an entropic quantity that is invariant under unitary transformations of its arguments, and the time evolution for both the process and its reverse is unitary.
An obvious choice of $\rho$ and $\tilde{\rho}$ is then to set $t=0$ so that the arguments of $D_1$ are the thermal states of the initial and final Hamiltonians.
By putting these into the definition of $D_1$ [Eq.~\eqref{eq:QuantumKL}], the free energy difference immediately comes out, and one arrives at Eq.~\eqref{eq:DissQWork}.

\citet{YungerHalpernGDV15_WC} show that a similar quantum equality\footnote{
In addition to this and the work fluctuation equation discussed in \cref{sec:WorkFluctuations},
 \citet{YungerHalpernGDV15_WC} also demonstrate that this equality holds for classical phase space fluctuations.
} holds relating the worst-case work with the order-$\infty$ R\'enyi divergence:
\begin{equation}
W_{\rm diss}^{\rm worst} = \kB T \rDiv{\infty}{\rho\left(t)}{\tilde{\rho}\right(\tau-t)}.
\end{equation}
The proof follows a similar path to that in \cite{ParrondoBK09} for $D_1$.
Noting the time-invariance of $D_\infty$, one plugs the thermal states corresponding to the beginning of each protocol into \eqref{eq:DinftyQuantum},
 to arrive at 
\begin{equation}
\rDiv{\infty}{\rho\left(0)}{\tilde{\rho}\right(0)}  = \ln \max_{i, j} \left\{ \exp\left[\beta\left(\tilde{E}_j - E_i - \Delta F\right)\right] \right\},
\end{equation}
where $\{\tilde{E}_j\}_j$ are the energies of the final Hamiltonian, and $\{E_i\}_i$ of the initial.
This quantity is clearly maximized when $\tilde{E}_j - E_i$ takes its largest value,  and so the entire right-hand side corresponds to $\beta W_{\rm diss}^{\rm worst}$.

A related equality for the case of finite-$\alpha$ R\'enyi relative entropies has been derived by~\citet{WeiP17}: 
\begin{equation}
\expt{\left(e^{-\beta W_{\rm diss}}\right)^\alpha} = \exp\left(-\alpha \beta \Delta F\right)  \exp\left( \rDiv{\alpha}{\rho(t)}{\tilde{\rho}(\tau - t)} \right).
\end{equation}
This is a generalization of the Jarzysnki equality, as seen by setting $\alpha=1$.
This equation opens up new experimental possibilities for determining R\'enyi divergences between quantum states (which is difficult to do via direct tomography).
This facilitates, for instance, the experimental testing of recent advances in modern thermodynamics that involve R\'enyi divergences between a state and its thermal equilibrium state~\cite{GourMNSYH15,BrandaoHNOW15,LostaglioJR15,LostaglioKDR15,CwiklinskiSHO15}.
Through the use of {\em single-qubit probes}~\cite{DornerJCPV12,DornerCHFGV13,Batalhao15,Xiong18} (a type of Ramsey interferometry -- see \cite{DeChiara18}), one can characterize the work-distribution,
 and then use the above equalities to determine the R\'enyi divergences between the initial and final thermal states~\cite{Guo17}.

We conclude by noting that recent developments in field of quantum fluctuation relations extend their remit beyond the regime of assumptions made by Tasaki~(e.g.\ \cite{AlhambraMOP16,Aberg18}).
From the success of the results presented thus far, 
 we suspect that wherever the \KL{} divergence might appears in these general contexts, an equivalent expression that uses R\'enyi divergences should also be possible.

\acknowledgments
The author is grateful for discussions, comments and suggestions from \mbox{Felix Binder}, \mbox{Oscar Dahlsten}, \mbox{Jayne Thompson}, \mbox{Vlatko Vedral}, and \mbox{Nicole Yunger Halpern}.
The author is financially supported by the Foundational Questions Institute ``Physics of the Observer'' large grant \mbox{FQXi-RFP-1614}.

\end{document}